\documentclass[twocolumn,showpacs,preprintnumbers,amsmath,amssymb]{revtex4}

\usepackage{graphicx}
\usepackage{dcolumn}
\usepackage{bm}

\begin{document}

\preprint{}

\title{ Comment on 
	"Scalar-tensor gravity coupled to a global monopole and flat rotation curves" by Lee and Lee. }

\author{P. Salucci}
 \altaffiliation{SISSA, Trieste.}
\email{salucci@sissa.it}
\author{G. Gentile}%
 \altaffiliation{SISSA, Trieste.}
 \email{gentile@sissa.it}
\affiliation{   SISSA, Trieste 
}%

\date{\today}

\begin{abstract}
 
The  recent paper by Lee and Lee (2004) may strongly leave the impression  that  astronomers have established  
that the  rotation curves of  spiral galaxies are flat.  We show  that the old  paradigm of  Flat Rotation Curves    lacks, today,   any  observational support and  following it at  face value  leads  to intrinsically flawed  alternatives to the Standard Dark Matter  Scenario.  On the other side, we claim that  the rich  systematics of   spiral galaxy rotation curves, that reveals, in the standard Newtonian Gravity framework,  the phenomenon of dark matter, in alternative scenarios,   works as a  unique   benchmark. 
\end{abstract}

\maketitle

\section{\protect\\ The Issue}

Lee and Lee (2004) consider a global monopole as a candidate for the galactic dark matter riddle and solve the Einstein Equations in the weak field and large $r$ approximations,
for  the case of Scalar Tensor Gravity (where $G=G_*(1+\alpha_0^2)$, with $G_* $ the bare Gravitational Constant). The potential of the triplet of the scalar field is written as: $V_M(\Phi^2)=\lambda/4  (\Phi^2-\eta^2)^2$,  the line element of the spherically symmetric static spacetime results as: $ds^2= -N(r)dt^2 + A(r) dr^2 + B(r) r^2 d\Omega^2$ where the functions  $N(r)$, $A(r)$,  $B(r)$ are given in their eq. 19. From the above,  
Lee and Lee (2004) write the geodesic equations, whose solution, for circular motions, reads:
$$
V^2(r) \simeq 8\pi G \eta^2 \alpha_0^2 +GM_\star(r)/r
\eqno(1)
$$

\begin{figure*}
\begin{center}
\includegraphics[ width=67mm,angle =-90]{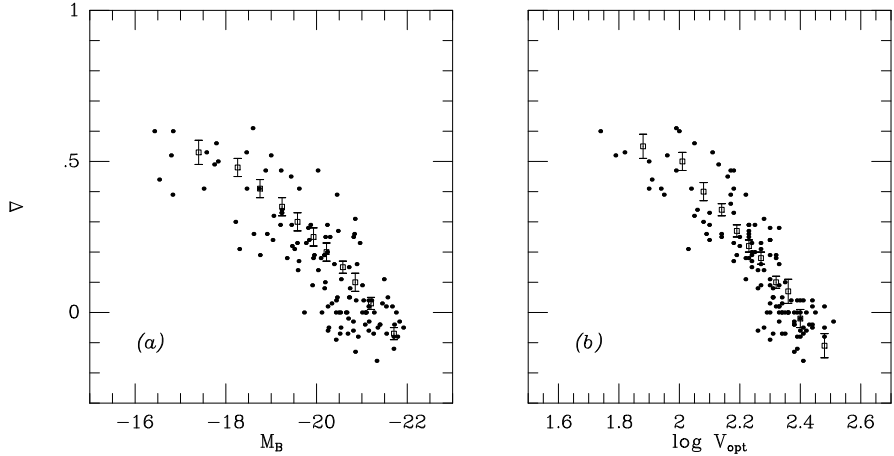}
\end{center}
\vskip -0.7truecm
\caption{Logarithmic gradient of the circular velocity $\nabla$ $vs.$  B absolute magnitude and 
$vs.$ $log \   V(R_{opt})$.
Lee and Lee (2004) predictions are  $\nabla(V_{opt})=0$ and $\nabla(M_B)=0$.}
\end{figure*}
where $M_\star (r)$ is the ordinary stellar mass distribution.
In  the above equation,   they interpret   the  first (constant)  term,  that  emerges   in addition to the 
standard Keplerian term, as the alleged constant (flat) value $V(\infty) $ that the circular velocities are thought to 
 asymptotically reach   in the external regions of galaxies, where  the (baryonic) matter 
contribution $GM_\star /r$  has   decreased from  the   peak  value by a factor of several. Furthermore,   they compare  the quantity  $  8\pi G \eta^2 \alpha_0^2$ with the spiral  circular velocities at  outer radii and   estimate: $\eta \sim 10^{17} GeV$.
The crucial features  of their theory (at the current stage)  are: the "DM phenomenon"  always  emerges  at  outer radii $r$  of  a galaxy  as a constant threshold    value below which the circular velocity $V(r)$ cannot decrease, regardless of  the distance between  $r$   and the location  of  the bulk  of  the stellar component.  

The  theory implies (or, at its present stage,  seems to imply)  the existence of an observational scenario in which  the rotation curves  of spirals are asymptotically flat  and   the new  extra-Newtonian (constant)  quantity  appearing in the modified  equation of motion,   can be  derived from the rotation curves themselves.   As a  result, the flatness of a RC becomes a   main  imprint for the Nature of the  "dark matter constituent".

The aim of this Comment is  to show   that   the above  "Paradigm  of Flat Rotation Curves" of spiral galaxies (FRC)
 has  no observational    support,  and to  present    its   
inconsistency by means of  factual evidence. Let us notice that we could have  listed a number of objects  with a serious  gravitating {\it vs.} luminous  mass discrepancy having  steep (and not flat) RC's and that only a minority of the observed
rotation curves can be considered as flat in the outer parts of spirals.  However, we think that    it is worth to discuss in  detail the phenomenology of the spirals' RC's,   in that we believe that it is the benchmark   of any  (traditional or   innovative)  work  on  "galactic dark matter",  including that of  Lee and Lee (2004).

The "Phenomenon of Dark Matter" was discovered in the late 70's (Bosma 1981, Rubin et al. 1980) as the  lack of the Keplerian fall-off   
in the circular velocity of spiral galaxies,   expected beyond their stellar  edges $R_{opt}$ (taken as 3.2 stellar disk exponential scale-lengths $R_D$).  In the  early years of the discovery  two facts  led to the concept of  Flat Rotation Curves: 

1) Large  part of  the evidence for DM was provided  by  extended,  low-resolution HI RC's of very luminous spirals (e.g. Bosma 1981) whose  velocity profile did show small radial variations.    
2) Highlighting  the few  truly flat  rotation curves was considered a way  to rule out the 
claim that  non Keplerian   velocity profiles originate from a  faint baryonic component distributed at large radii.   

It was soon realized that HI  RC's of   high-resolution and/or of galaxies of  low luminosity  did  
vary with radius, that baryonic (dark) matter was  not a plausible candidate for the cosmological DM, and   finally,  the prevailing   Cosmological Scenario (Cold Dark Matter)  did predict  galaxy halos  with  rising as well as with  declining rotation curves (Navarro, Frenk and White, 1996).
The FRC paradigm was  dismissed by researchers  in  galaxy kinematics in the early 90's (Persic et al.  1988, Ashman, 1992), and  later  by  cosmologists (e.g. Weinberg, 1997). Today,  the  structure of the DM  halos and their rotation speeds  is  thought to have   a central role  in Cosmology  and  a strong link to  Elementary Particles  via the Nature of their constituents, (e.g.  Olive 2005)  
and a  careful interpretation  of the   spirals' RC's is considered crucial.

\section{\protect\\ The Observational  Scenario }

Let us stress that a  FRC   is not  a proof of  the existence of  dark matter in a galaxy. In fact, the circular 
velocity due to a  Freeman stellar disk   has a   flattish  profile between 2 and 3 disk scale-lengths.   
Instead,  the evidence  in  spirals of   a serious mass discrepancy, that   we  
 interpret  as the effect of     a dark  halo enveloping  the stellar disk,  originates from   the fact  that,  in their optical regions, the   RC are often  steeply   rising.

Let us  quantify the above statement   by plotting   the average value of the   RC logarithmic slope,   $\nabla \equiv  \ (dlog \  V / dlog \  R)$  between two and three  disk scale-lengths as a function of the rotation speed  $V_{opt}$  at the disk edge $R_{opt}$.
  We remind that, at 3 $R_D$  in  the case  of no-DM  self-gravitating  Freeman disk, $\nabla =-0.27$
 in any object,   and that   in  the  Lee and Lee proposal $\nabla  \sim 0 $ (see eq. 1).  

 We consider the sample of 130  individual  and 1000 coadded RC's  of normal spirals, presented in Persic, Salucci \& Stel (1996) (PSS). We  find (see Fig. 1b):
$$
\nabla = 0.10-1.35 \ log {V_{opt}\over {200~ km/s}} 
\eqno(2a)
$$
(r.m.s. = 0.1),  where $80\  km/s  \leq   V_{opt}\leq 300  \    km/s$.
A similarly tight relation links $\nabla$ with the galaxy absolute magnitude (see Fig. 1a). 
For dwarfs,  with $40 \ km/s  \leq   V_{opt}\leq 100 \ km/s $,  we take the results by  Swaters (1999): 
$$
\nabla = 0.25-1.4 \ log {V_{opt}\over {100\  km/s}}
\eqno(2b)
$$
(r.m.s. = 0.2) that results   in good agreement with the extrapolation of eq. 2. 
The {\it large range}  in  $\nabla$  and the high   values of these quantities, 
implied by  eq.  2 and evident in Fig. 1,  are  confirmed by  other studies of independent  samples   (e.g.  Courteau 1997, see Fig. 14 and  Vogt et al. 2004, see figures inside).
  
Therefore, in disk systems, in region where the stars reside,  the RC slope takes  values in the range:
 $$
-0.2 \leq \nabla \leq 1
 $$
 i.e. it   covers  most of the range that a circular velocity slope  could take (-0.5 (Keplerian) , 1 (solid body)).
Let us notice that the difference between the  RC slopes  and the no-DM  case is almost as 
severe  as the difference between the former  and the alleged value of  zero.

It is apparent  that  only a  very minor  fraction of RC's can be 
considered as   flat. Its rough estimate  can be derived in simple way. At  luminosities $L<L_*$, 
($L_*=10^{10.4}\  L_{B\odot}$ is the knee of the Luminosity Function  in the B-band) the 
spiral Luminosity Function can be assumed  as   a  power law:  $\phi(L) dL  \propto L^{-1.35} dL$, then, by means of  the  Tully-Fisher relationship $L/L_* \simeq  (V_{opt}/(200~ km/s))^3$ (Giovanelli et al., 1997) combined with  eq. 2a, one gets:
$
n(\nabla) d\nabla \propto 10^{0.74 \nabla} d\nabla
$ finding  that the objects with a  solid-body RC ($0.7  \leq \nabla  \leq 1$) are  one order of magnitude more numerous  than those with a "flat" RC ($-0.1 \leq \nabla  \leq 0.1$). 
In short,  there is plenty of evidence of galaxies whose inner regions show a  very steep RC,   that  in the Newtonian + Dark Matter Halos framework, implies that they are dominated by  a dark component,  with a density profile  much shallower than the "canonical"  $r^{-2}$  one.  
 \begin{figure}
\vskip 1cm
\begin{center}
\includegraphics[width=49mm]{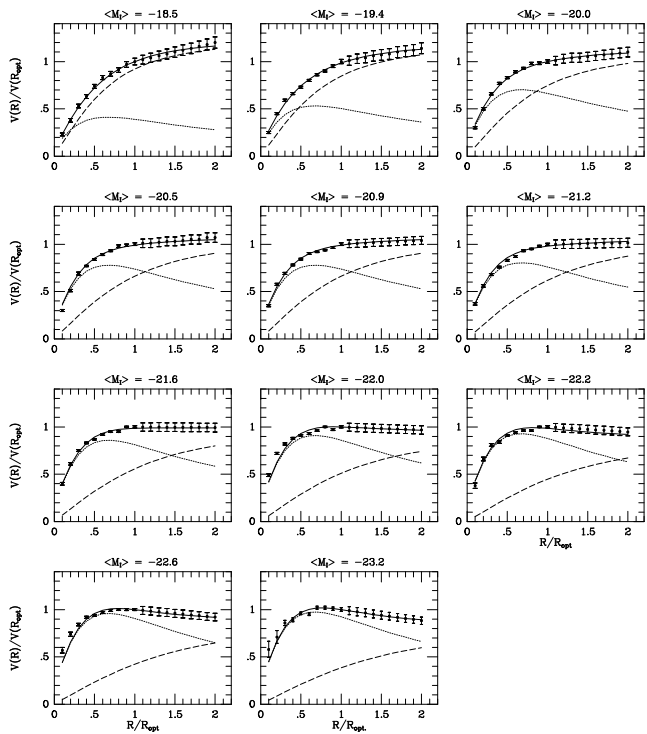}
\end{center}
\vskip -0.3truecm
\caption{ The Universal Rotation Curve }
\end{figure}

At outer radii (between 6-10 disk scale-lengths)  the observational data are obviously more scanty,
 however,  we observe a varied and systematics zoo of rising, flat, and declining RC's profiles
 (Gentile et al. 2004;  Donato et al. 2004).
 
\section{\protect\\ Discussion }

The evidence from about 2000 RC's of normal and dwarf  spirals unambiguously shows  the existence of a  systematics in the 
rotation curve profiles inconsistent with the  Flat Rotation Curve paradigm.  The non stellar  term in eq. 1 must have a radial dependence in each galaxy  and vary among galaxies.  To show this let us  summarize the RC systematics.  
In general, a  rotation curve of a spiral, out to 6 disk scale-lengths,  is well described by the following function:
$$
V(x)=V_{opt} \biggl[ \beta  {1.97x^{1.22}\over{(x^2+0.782)^{1.43}}} + (1-\beta)(1+a^2)\frac{x^2}{x^2+a^2} \biggl]
$$
where $x \equiv R/R_{opt}$ is the normalized radius, $V_{opt}=V(R_{opt})$, $\beta=V_d^2/V_{opt}^2$, $a=R_{core}/R_{opt}$ are free parameters, $V_d$ is
the contribution of the stellar disk at $R_{opt}$ and $R_{core}$ is the core
radius of the dark matter distribution.
Using a sample of $\sim$ 1000 galaxies, PSS found that,
out to the farthest radii with  available data, i.e.  out to $6\ R_D$, the luminosity specifies the above free parameters 
i.e. the main average properties of  the axisymmetric rotation field of spirals and, therefore,  of  the  related mass distribution. In detail,  eq. 2 becomes the expression for the 
{\it Universal Rotation Curve} (URC, see Fig. 2 and PSS for important details). Thus,  for a  galaxy of luminosity $L/L_*$ (B-band) and normalized radius $x$ we have (see also Rhee,  1996):  
$$
V_{URC}(x) =V_{opt} \biggl[ \biggl(0.72+0.44\,{\rm log} {L \over
L_*}\biggr) {1.97x^{1.22}\over{(x^2+0.782)^{1.43}}} +
$$
\vskip -0.85truecm
$$ 
   \biggl(0.28 - 0.44\, {\rm log} {L \over L_*}
\biggr)  \biggl[1+2.25\,\bigg({L\over L_*}\biggr)^{0.4}\biggr]  { x^2 \over
x^2+2.25\,({L\over L_*})^{0.4} } \biggr]^{1/2} 
$$
The above can be written as:  
$V^2(x)=G (k M_\star/x+ M_h(1) F(x,L))
$
where $M_h(1)$ is the halo mass inside $R_{opt}$ and  k  of the order of unity. 
Then, differently from the Lee and Lee (2004) claim and the FRC paradigm,  the "dark" contribution $F(x,L)$ to the RC  varies with radius, namely as  $x^2/(x^2+a^2)$, $a =const$ in each object.  Finally, also  the extrapolated  "asymptotic  amplitude $V(\infty)$" varies,  according to the galaxy luminosity, between $50\  km/s $ and $ 250\ km/s $ (see also PSS)   in disagreement   with the  Lee and Lee (2004)  predicted constant value of $  8\pi G \eta^2 \alpha_0^2 \sim 300\  km/s $.

Let us conclude with an  {\it important} point:  this paper is not
intended to discourage testing  out whether a theory, alternative to the
DM paradigm,  can account for an outer  flat rotation curve, but to make
us sure that  this is  the  (simplest)  first step  of a project meant
to  account for  the  actual complex  phenomenology of
rotation curves of spirals and of the implied physical relevance of the  mass discrepancy (e.g. Gentile et al. 2004).

{}
\end{document}